\documentclass{iau} 
\usepackage{graphicx}
\usepackage{xcolor}
\usepackage[hyphens]{url}
\usepackage{hyperref}
\hypersetup{
     colorlinks  = true,
     linkcolor   = blue, 
     anchorcolor = BrickRed,
     citecolor   = blue,
     filecolor   = blue,
     menucolor   = blue,
     runcolor    = cyan,
     urlcolor    = blue
}
\usepackage[authoryear]{natbib}

\pubyear{2019}
\volume{355} 
\jname{The Realm of the Low-Surface-Brightness Universe}
\editors{D. Valls-Gabaud, I. Trujillo \& S. Okamoto, eds.}

\def\etal{\textit{et al.}}

\title[Energy production over all time] 
{Measuring energy production in the Universe over all wavelengths and all time}

\author[Simon P. Driver]   
{Simon P. Driver}

\affiliation{
International Centre for Radio Astronomy Research (ICRAR), University of Western Australia, 35 Stirling Highway, Crawley, Perth, WA6009, Australia}

\begin{document}

\maketitle

\begin{abstract}
The study of the extragalactic background light (EBL) is undergoing a
renaissance. New results from very high energy experiments and deep
space missions have broken the deadlock between the contradictory
measurements in the optical and near-IR arising from direct versus
discrete source estimates. We are also seeing advances in our ability
to model the EBL from $\gamma$-ray to radio wavelengths with improved
dust models and AGN handling. With the advent of deep and wide
spectroscopic and photometric redshift surveys, we can now subdivide
the EBL into redshift intervals. This allows for the recovery of the
Cosmic Spectral Energy Distribution (CSED), or emissivity of a
representative portion of the Universe, at any time. With new
facilities coming online, and more unified studies underway from
$\gamma$-ray to radio wavelengths, it will soon be possible to measure
the EBL to within 1 per cent accuracy. At this level correct modelling
of reionisation, awareness of missing populations or light, radiation
from the intra-cluster and halo gas, and any signal from decaying
dark-matter all become important. In due course, the goal is to
measure and explain the origin of all photons incident on the Earth's
surface from the extragalactic domain, and within which is encoded the
entire history of energy production in our Universe.
\end{abstract}

\firstsection 

\begin{figure}[b]
\begin{center}
 \includegraphics[width=\textwidth]{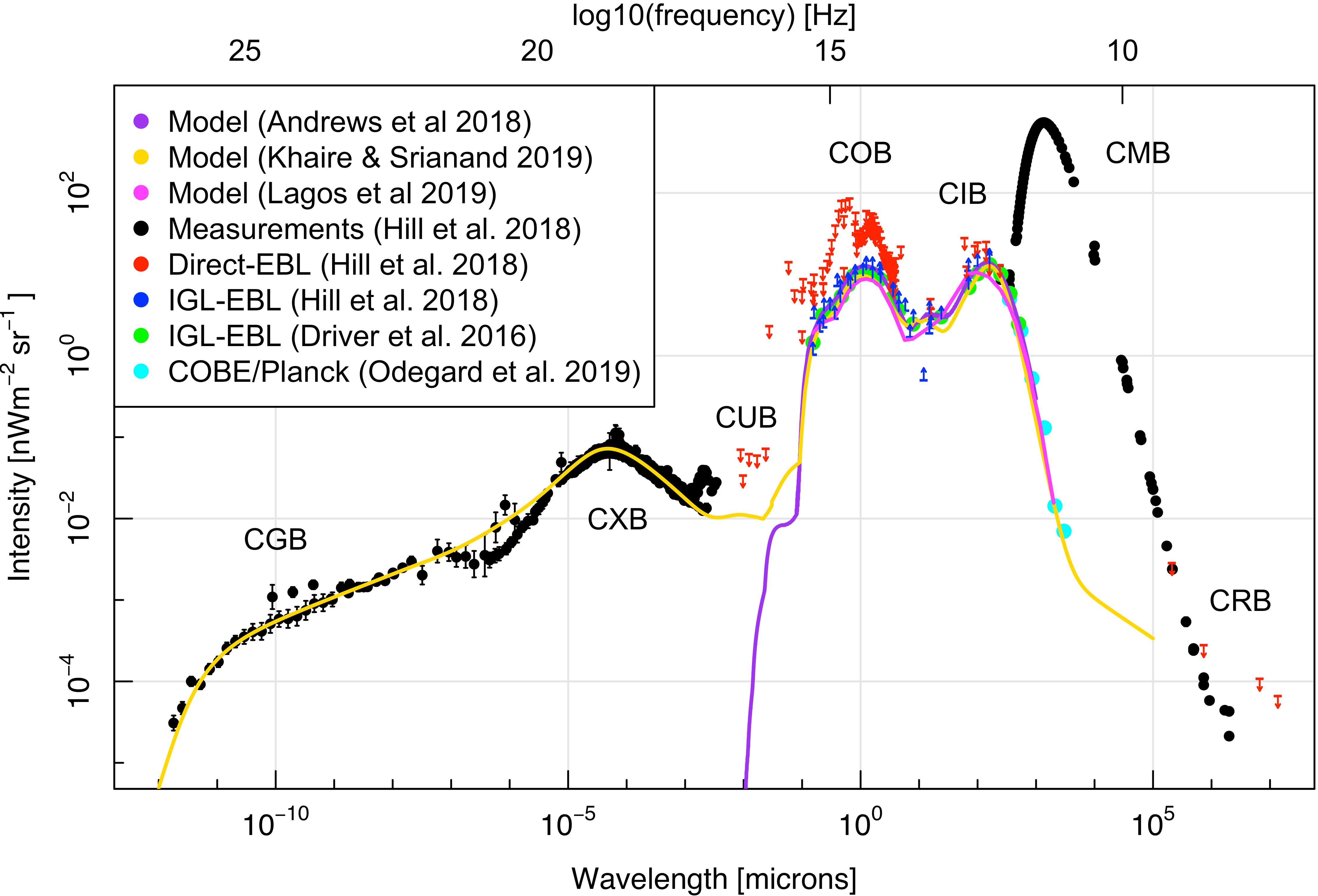} 
 \caption{A compendium of recent EBL measurements, mainly based on
   data assembled by \cite{hill2018} and also including the
   \cite{andrews2018} UV-far-IR model (purple line), the recent
   semi-analytic Shark model \citep[][, in magenta]{lagos2019} and the
   \cite{ks19} $\gamma$-ray to radio model (gold line).}
  \label{fig:ebl}
\end{center}
\end{figure}

\section{Introduction}
The extra-galactic background light \citep[EBL, see review
  by][]{mattila2019}, represents the radiation incident from a
steradian of extragalactic sky onto a square metre of the Earth's
atmosphere, i.e., the photon-flux that originates from outside our
Galaxy. Encoded in the EBL is the entirety of photon production
pathways that have existed from the Big Bang to the present day. This
is a remarkable concept, as to measure and model the EBL, is to
explain all photon-energy production over all time.

Fig.~\ref{fig:ebl} shows a recent compendium of EBL measurements,
primarily taken from \cite{hill2018}, but augmented with some
additional data from COBE/Planck \citep{odegard2019}, the recent UV to
far-IR model from \cite{andrews2018} (purple line), a recent
semi-analytic model from Shark \citep{lagos2019}, and the more
extended panchromatic model from \cite{ks19} (solid gold line). This
overall EBL can be subdivided into a number of ``backgrounds'', as
indicated, each spanning specific wavelength ranges, with the
distinction driven mostly by the technologies required to obtain the
measurements.

The most notable contributor is the Cosmic Microwave Background (CMB),
arguably not actually part of the EBL, which dominates the incident
flux in terms of photon number and photon energy. The CMB sits slightly
apart from the other backgrounds, in that it originates from the
energy remains of the early Big Bang era leading up to decoupling. The
remainder of the EBL is made from photon-production mechanisms which
have occurred {\it since} decoupling. 

At about 10 per cent of the CMB, are the Cosmic Optical Background
(COB) and the Cosmic Infrared Background (CIB). There are three key
processes at play here \citep[see for example][]{gilmore2012, andrews2018, ks19}: star-formation; accretion of
material onto black holes (predominantly AGN activity); and dust
reprocessing in which approximately half of the photons generated in
the former two phases are attenuated by dust grains within the host
system and reradiated in the far-IR \citep{dunne2003,driver2008}.

The Cosmic Ultraviolet Background (CUB) and Cosmic Radio Background
(CRB) are less well studied \citep[see][for a summary of both
  backgrounds]{hill2018}, and represent fertile ground for future
investigation and modelling \citep[see in particular discussion of the
  CUB in][]{ks19}. Currently these are primarily constrained by
observations from the Extreme Ultraviolet Explorer (EUV), and a
relatively small number of radio surveys.

Finally, we have the high-energy Cosmic $\gamma$-ray background
\cite[CGB; see for example][]{ajello2015} and the Cosmic X-ray
Background \citep[CXB; see for example][]{cappelluti2017}. These are
fairly well constrained and comprised of photons originating from
high-energy processes such as x-ray binaries, AGN, cooling of gas,
supernovae, and shock events: essentially plasma astrophysics. And a
rich area for placing constraints on decaying dark matter models,
for example \cite{darkmatter}.

Because of the energy dominance of the COB and CIB these are, after
the CMB, the most studied wavelength regions and the main focus of
this article. However, before moving on, it is worth advocating the
need for further work in particular soft x-ray, extreme ultraviolet,
radio, and very high ($>TeV$) scales. All of these windows, except for
the EUV \citep[see][]{cooray2016}, are being extended through new
facilities, e.g., the upcoming Cherenkov Telescope Array, the recently
launched eROSITA satellite, and higher sensitivity wide and deep radio
surveys at short (ASKAP, MeerKAT), and long wavelengths (LOFAR,
MWA). The extreme-UV however looks destined to remain fairly uncharted
territory, with no obvious plans for a space mission to cover this
wavelength region (although we note the proposed {\it Messier}
mission --- see these proceedings --- will extend down to 0.15micron). One
motivation to push further into the EUV might be the recent study of
\cite{mattila2017a} who used dark cloud observations to detect an
anomalous UV photon-flux increasing into the UV
\citep{mattila2017b}: could this be an interesting future window for
decaying dark-matter searches?

\begin{figure}[h]
\begin{center}
 \includegraphics[width=\textwidth]{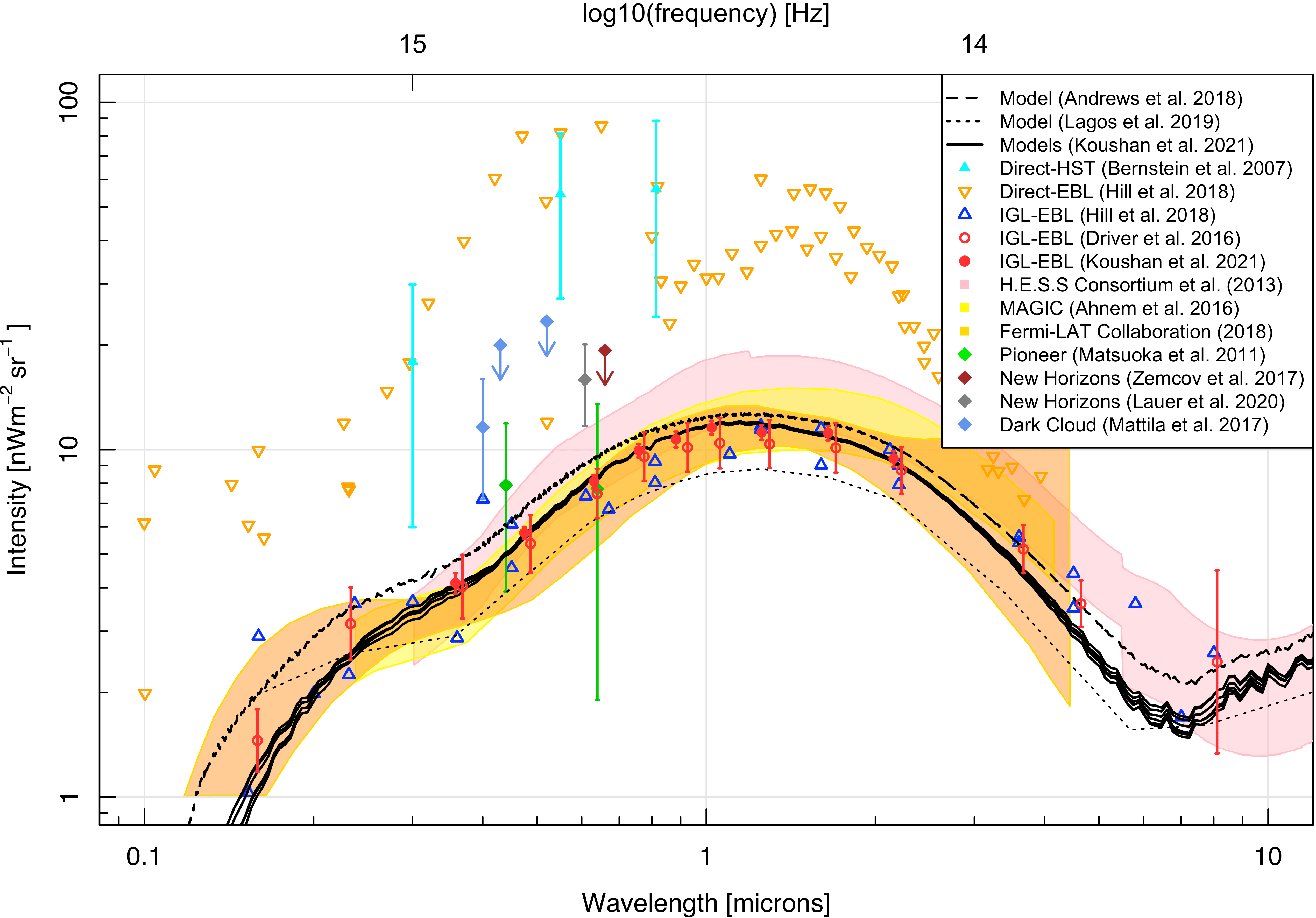} 
 \caption{A zoom in of the COB and CIB region from Fig.~\ref{fig:ebl},
   but including data from the Very High Energy experiments and deep
   space probes Pioneer 10/11 and New Horizons. These seem to
   corroborate the IGL-EBL measurements.}
  \label{fig:eblzoom}
\end{center}
\end{figure}

\section{The optical controversy}
Most obvious from Fig.~\ref{fig:ebl} is a significant disparity in the
COB and to some extent the CIB. Fig.~\ref{fig:eblzoom} shows a zoom in of
this region which now includes the errorbars associated with the
measurements. These estimates predominantly fall into two camps:
Observations from direct background measurements (direct-EBL); and
observations from integrated galaxy counts (IGL-EBL). The former
method should capture {\it all} the photon flux, whereas the latter
only captures the photon-flux from discrete detectable sources, i.e.,
galaxies and quasars. The fact that the direct-EBL measurements can be
a factor of 4---10$\times$ above the IGL-EBL, can have a number of possible
explanations. The least exciting is that one method is simply in
error. The most exciting is that there are significant unknown
photon-production pathways occurring outside of the detectable galaxy
population. Possibilities for the latter might include missing
populations of diffuse galaxies, excessive stripped gas,
photon-production from a diffuse component of the IGM via some unknown
process, or more exotic possibilities such as decaying dark-matter or
dark-mater/matter interactions. Regardless, it is clearly important to
understand the nature of this discrepancy.

\begin{figure}[h]
\begin{center}
 \includegraphics[width=\textwidth]{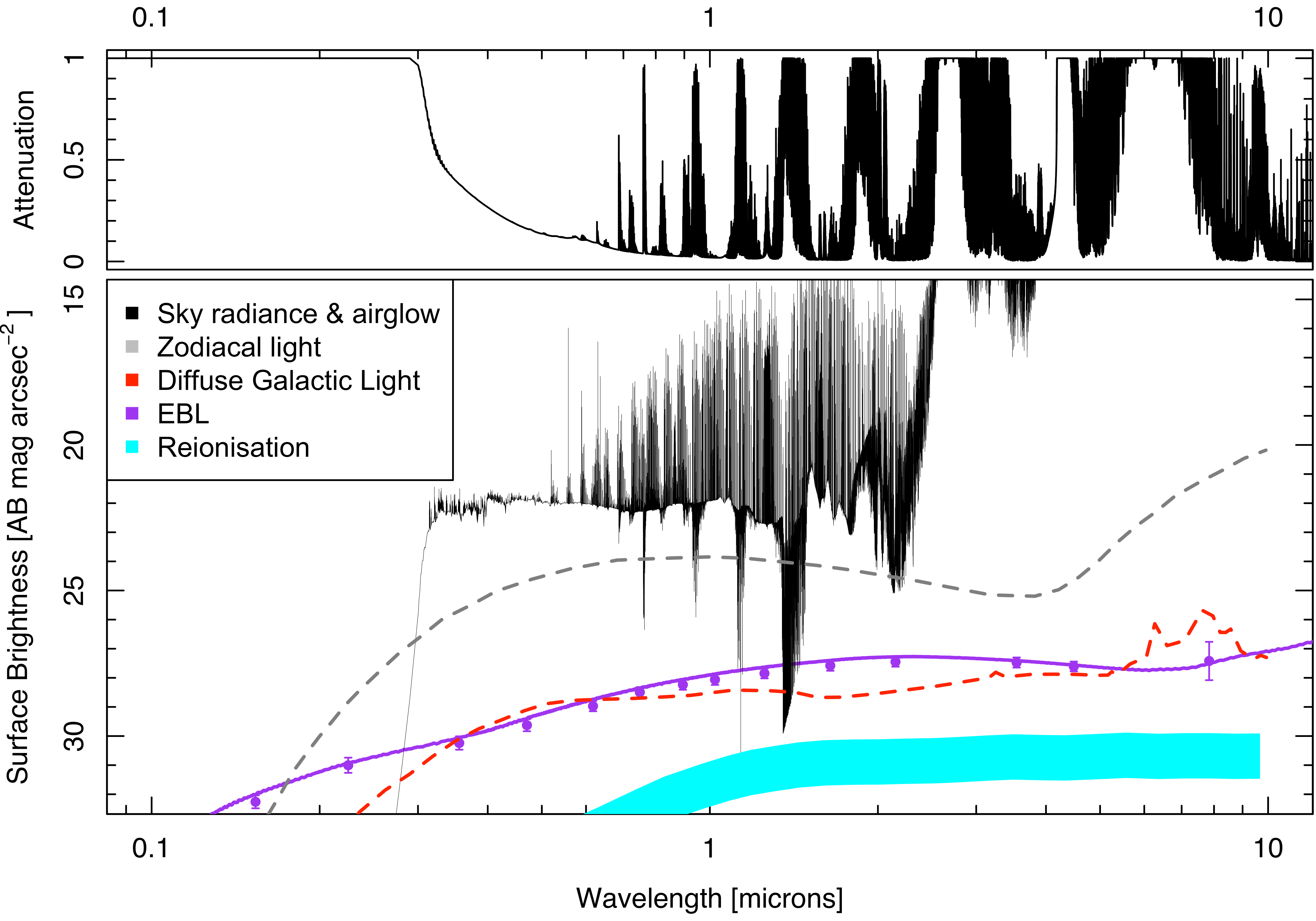}
  \caption{({\it main panel}) The spectral energy distributions of
    various backgrounds including, the night sky, zodiacal light, the
    extragalactic background light, the diffuse Galactic light, and
    the expected signal of reionisation. The figure is shown in
    atypical units of Surface Brightness in AB mag/arcsec$^{2}$. ({\it
      top panel}) The attenuation of external radiation by the Earth's atmosphere.}
  \label{fig:back}
\end{center}
\end{figure}

Direct-EBL estimates are typically taken from measurements of the
absolute sky background from well calibrated detectors, most notably
data from the {\it Hubble Space Telescope}, which sits above the
Earth's atmosphere. However, this still requires further background
subtraction of the Zodiacal Light (Zodi) and any Diffuse Galactic
Light (DGL). Fig.~\ref{fig:back} highlights this issue by showing the
flux of various components including that from the Earth's night sky,
the Zodi (dependent on Ecliptic coordinates), the Diffuse Galactic
Light (dependent on Galactic coordinates), the EBL (isotropic), and
the component of the EBL from reionisation (isotropic but with a
highly uncertain shape and amplitude). Very roughly the Zodi is about
10 per cent of a dark site moon-less night sky, the EBL 1-10 per cent
of the Zodi, the EBL and DGL comparable (with the latter highly dependent on
direction), and the reionisation signal 1 per cent of the EBL
\citep[see][for more details]{zemcov2011,cooray2012}. Hence the
near-IR reionisation signal is around a thousandth of a per cent of
the Earth's night sky flux. Moreover, a 1 per cent systematic error in
the subtraction of the Zodi can lead to a 100 per cent error in direct
measurements of the EBL \citep[a point well made
  by][]{berstein2007}. HST, of course sits mostly above the atmosphere
but at the start and end of an observation the telescope's attitude
can encroach on the Earth's limb introducing a component of
Earth-shine (i.e., some additional small fraction of the night-sky
spectrum).

\begin{figure}[h]
\begin{center}
 \includegraphics[width=\textwidth]{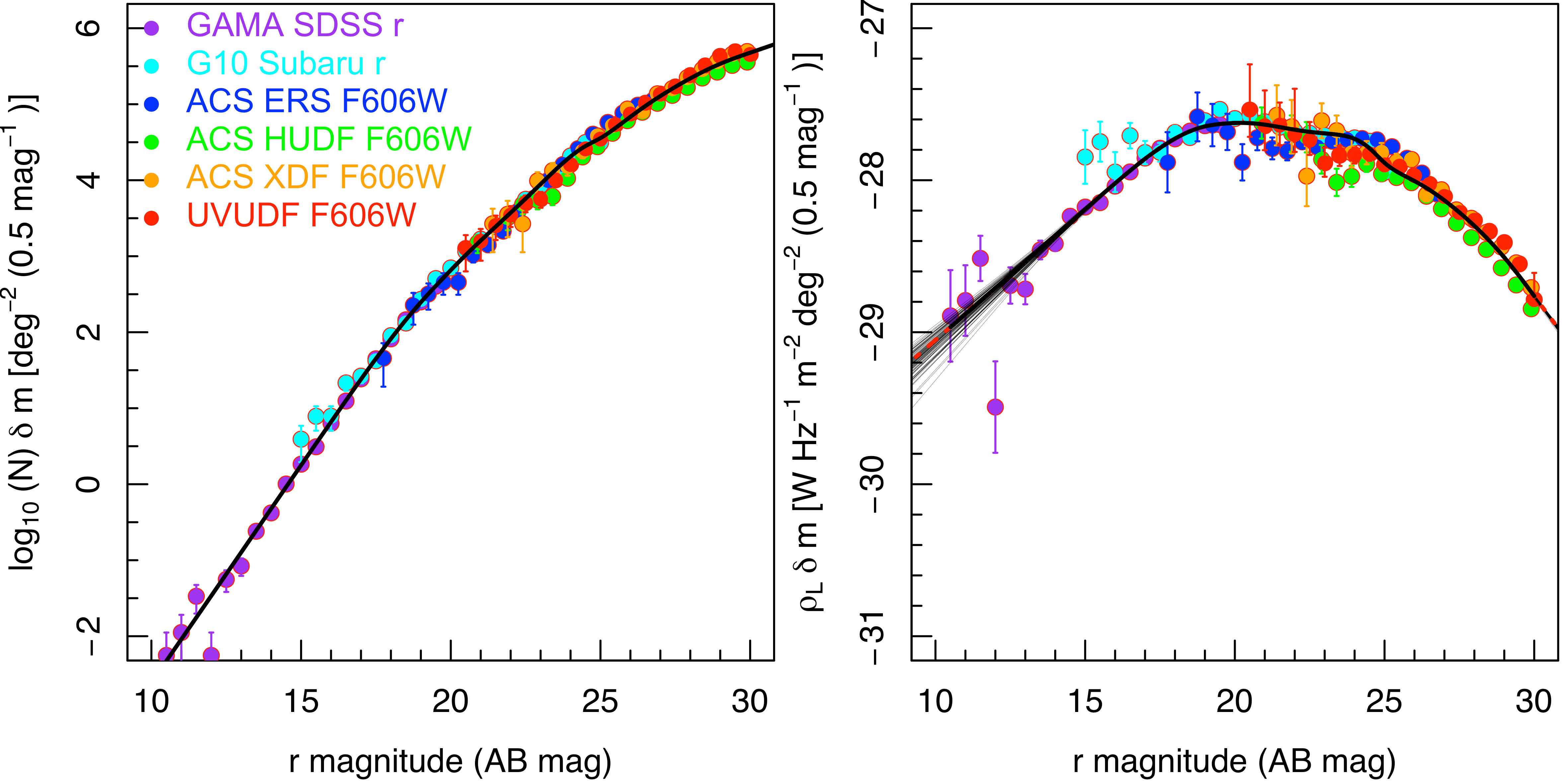} 
 \caption{The IGL method relies on assembling galaxy number-counts
   ({\it left panel}) and then integrating the flux contribution of
   each magnitude interval ({\it right panel}). Note that in most
   optical/near-IR bands the contribution is well bounded with minimal
   extrapolation required, and that the peak contribution is at
   relatively intermediate magnitudes. Improvements will come not from
   deeper or wide data but improved methodologies in flux measurements
   and sample sizes at the mid-mag range.}
  \label{fig:counts}
\end{center}
\end{figure}

The IGL-EBL estimates come from combining galaxy number-counts from
the widest and deepest imaging across multiple wavebands. This also
requires some extrapolation at very faint and very bright magnitudes
\citep[see for example][]{driver2016}. Recently, with advances in very
wide (e.g., GALEX, SDSS, VST KiDS, WISE etc), and very deep surveys
(e.g., HST Candles etc), most optical and near-IR bands have
number-count gradients which can be seen to gradually flatten from the
canonical bright non-expanding Euclidean expectation of $\delta
\log_{10} N(m) / \delta m = 0.6$ to below $0.4$, the gradient of
maximum contribution to the overall luminosity density. Assuming the
counts continue to decline in a monotonic fashion the IGL-EBL is
therefore bounded and measurable with a relatively small extrapolation
error. Fig.~\ref{fig:counts} highlights how an IGL-EBL constraint is
made for a particular waveband. The left-panel shows the raw galaxy
number-counts from a number of surveys, and the right-panel shows the
contribution of each magnitude interval to the overall IGL-EBL
measurement. The IGL-EBL is recovered by integrating under the
right-panel curve with a model-fit, or an extrapolated spline-fit
\citep[see for example][]{driver2016}. The concern with the IGL-EBL
method, is whether surveys are missing populations of very diffuse
galaxies leading to an underestimation of the number-counts. The
relatively good agreement between number-count data from so many
heterogeneous depth datasets would suggest this is not the case, but
nevertheless the issue does remain a valid concern at some level.

\section{Very High Energy and deep space missions to the rescue}
Recently, the disparity in Fig.~\ref{fig:eblzoom} appears to at least
be partially resolved from the inclusion of two new lines of argument:
Very High Energy (VHE) constraints, and measurements from the Pioneer
10 \& 11 \citep{matsuoka2011} and New Horizons \citep{zemcov2017,lauer2021} deep
space probes which are sufficiently far from the Sun for a much
diminished Zodi background \citep{bock2012,cooray2016}.

VHE studies rely on the interaction of the expected power-law
distribution of GeV photons from distant Blazars preferentially
interacting with $\sim 1$micron photons within the EBL through the
production of $e^+e^-$-pairs. This integrated interaction along the
pathlength of the TeV radiation results in a decrement in the received
Blazar spectrum at energies where the interaction is expected to be
maximum. This method requires the adoption of a COB model, and the
VHE data constrains the amplitude of the model. Recently the
H.E.S.S. and MAGIC teams reported the need for a small upward
normalisation over the IGL-EBL data of just 20 per cent, with about a
20 per cent uncertainty, i.e., essentially consistent. More recently
the largest study to date from the Fermi-LAT collaboration of $\sim
750$ Blazars reported full consistency with the IGL data
\citep{fermilat}. The VHE results are shown as the three solid bands
on Fig.~\ref{fig:eblzoom}. Current efforts are also underway to
constrain not only the normalisation, but the actual COB spectral
energy distribution shape. This is significantly harder but starting
to place useful constraints on the COB SED, see for example
\cite{biteau2015} and \cite{veritas2019}.

The Pioneer 10/11 and New Horizons deep space probes, both contain
relatively stable imaging cameras and these cameras which can be used
for direct-EBL measurements
\citep[see][]{matsuoka2011,zemcov2017,lauer2021}. As Zodiacal light
strongly drops as one moves out in the Solar System
\citep[see][]{bock2012,cooray2016}, its overwhelming impact is
diminished and its accurate subtraction less
problematic. Unfortunately the field-of-view of the onboard cameras
are small and sensitivity poor. One cannot also not entirely rule out
some degradation of the system throughput and response functions with
time. Nevertheless the three studies to data show a consistent picture
and agree more closely with the IGL-EBL data, albeit with fairly large
errorbars (see Fig.~\ref{fig:eblzoom}. Note that \cite{koushan2021}
have now decreased the uncertainty on the EBL-IGL measurements to $<5$ per cent.

\section{Entering the era of precision EBL studies}
The recent VHE and deep space data appear to provide compelling
evidence that the IGL-EBL measurements pretty much represents the full
COB to within $\sim$20 per cent \citep{driver2016}, with little room
for significant photon-flux from new populations, diffuse light,
decaying dark matter or any other furphy, at least arising in optical
wavebands. To first order this looks to confirm that our understanding of
the amount of star-formation that has occurred in the Universe over
all time is correct to within around 20 per cent, i.e., the current
level of error and scatter in the IGL, VHE, and deep space
constraints. A 20 per cent uncertainty, however, is still significant
and could easily mask a {\it modest} population of diffuse low surface
brightness galaxies, or other photon-production pathways. This then
motivates the reduction of the errors through direct-EBL, VHE and
IGL-EBL methods to somewhere around the 1 per cent level. A goal which
would represent a remarkable empirical feet but also entirely
attainable with the next decade.


\section{The SkySURF program}
While the discrepancy in the HST direct-EBL estimates and the IGL-EBL
data appears to be resolved, the explanation for the discrepancy is
still not known but cannot be extragalactic. There are a number of
obvious possibilities. The first is a limited understanding of the
Zodiacal Light in the inner Solar System, which in turn implies a
limit in our understanding of the Solar System dust distribution and
properties. The second is an additional source of contamination in HST
data, plausibly a component of Earth-shine given HST's relatively
low-orbit and the tendency to pack orbits close to the Earth-limb. A
more radical and exciting prospect might be additional optical
radiation emanating from the Galactic Halo, or a brighter than expected
contribution from the DGL.

Led by Prof Rogier Windhorst and Dr Rolf Jansen at Arizona State
University, a team of US and Australian scientists are looking to
address this by reprocessing the entire HST Advanced Camera for
Surveys and Wide-field Camera 3 archive, as part of an HST Cycle
27---29 SkySURF Archival program (AR-15810). The goals are to conduct
both direct-EBL sky measurements, as well as obtain refined
medium/deep galaxy number-counts using our new source finding code
ProFound \citep{robotham2018}, to also improve the IGL-EBL constraints.

The direct measurements, in particular, will be used to untangle the
three distinct backgrounds: the Zodi, the DGL, and the EBL; by using
their distinct positional dependencies: Ecliptic, Galactic, and
isotropic respectively.

The HST number-count measurements will be combined with similar work
underway for the European Southern Observatory's Visible Survey
Telescope (VST) and Visible Infrared Survey Telescope (VISTA)
kilo-degree surveys (KiDS and VIKING respectively). Combined VST,
VISTA and HST data will therefore provide a homogeneously defined set
of galaxy number-counts from $u$ to $K$ and from AB 10$^{th}$ to
30$^{th}$ magnitude. However, significant systematics need to be
overcome related to issues such as star-galaxy separation, galaxy
fragmentation, over-blending, and the aforementioned sensitivity to
diffuse populations.


\begin{figure}[h]
\begin{center}
 \includegraphics[width=\textwidth]{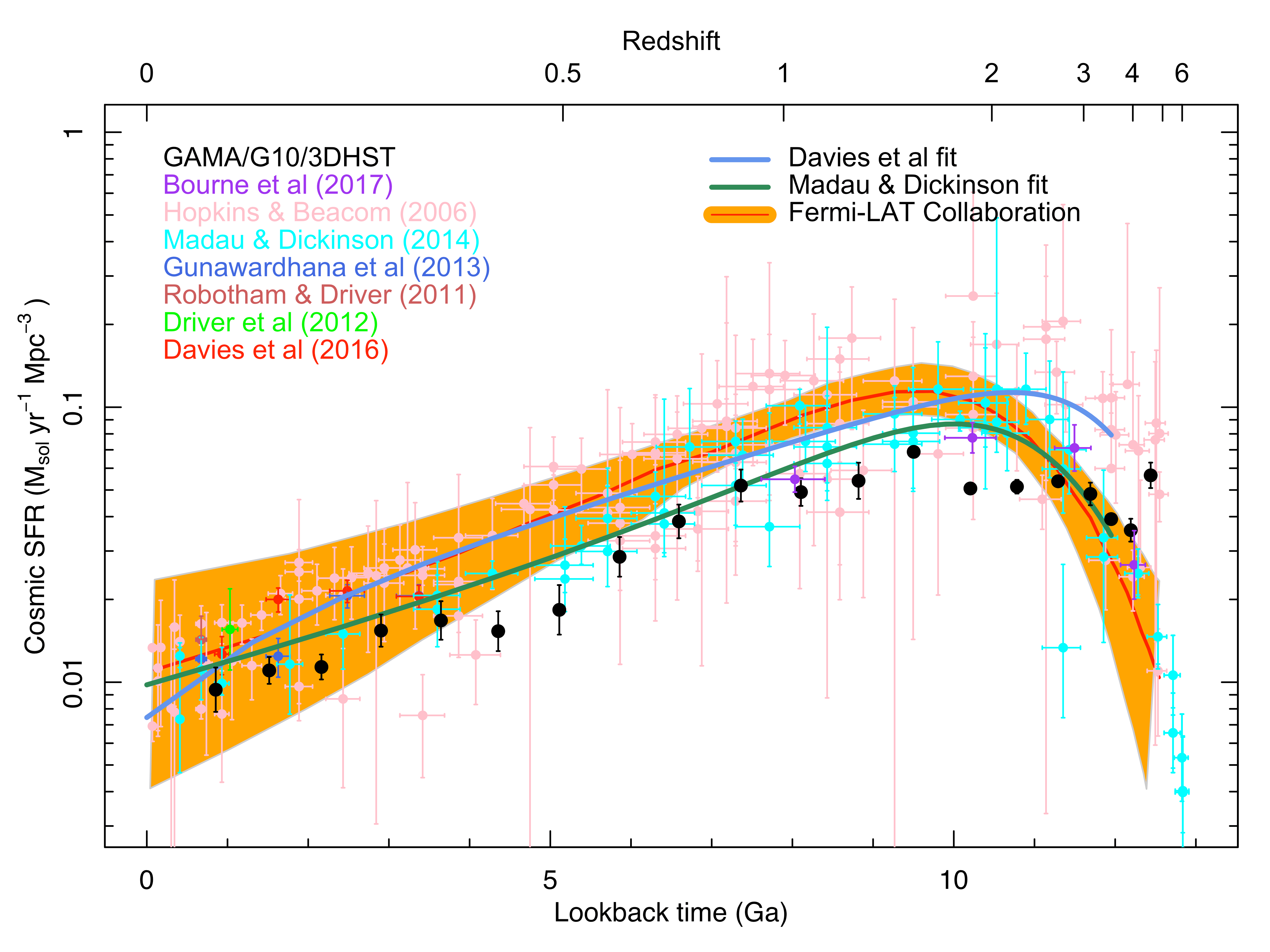} 
 \caption{The latest cosmic star-formation history plot used as the
   starting point for our model and including the recent VHE
   constraints from the \cite{fermilat}.}
  \label{fig:csfh}
\end{center}
\end{figure}

\begin{figure}[h]
\begin{center}
 \includegraphics[width=\textwidth]{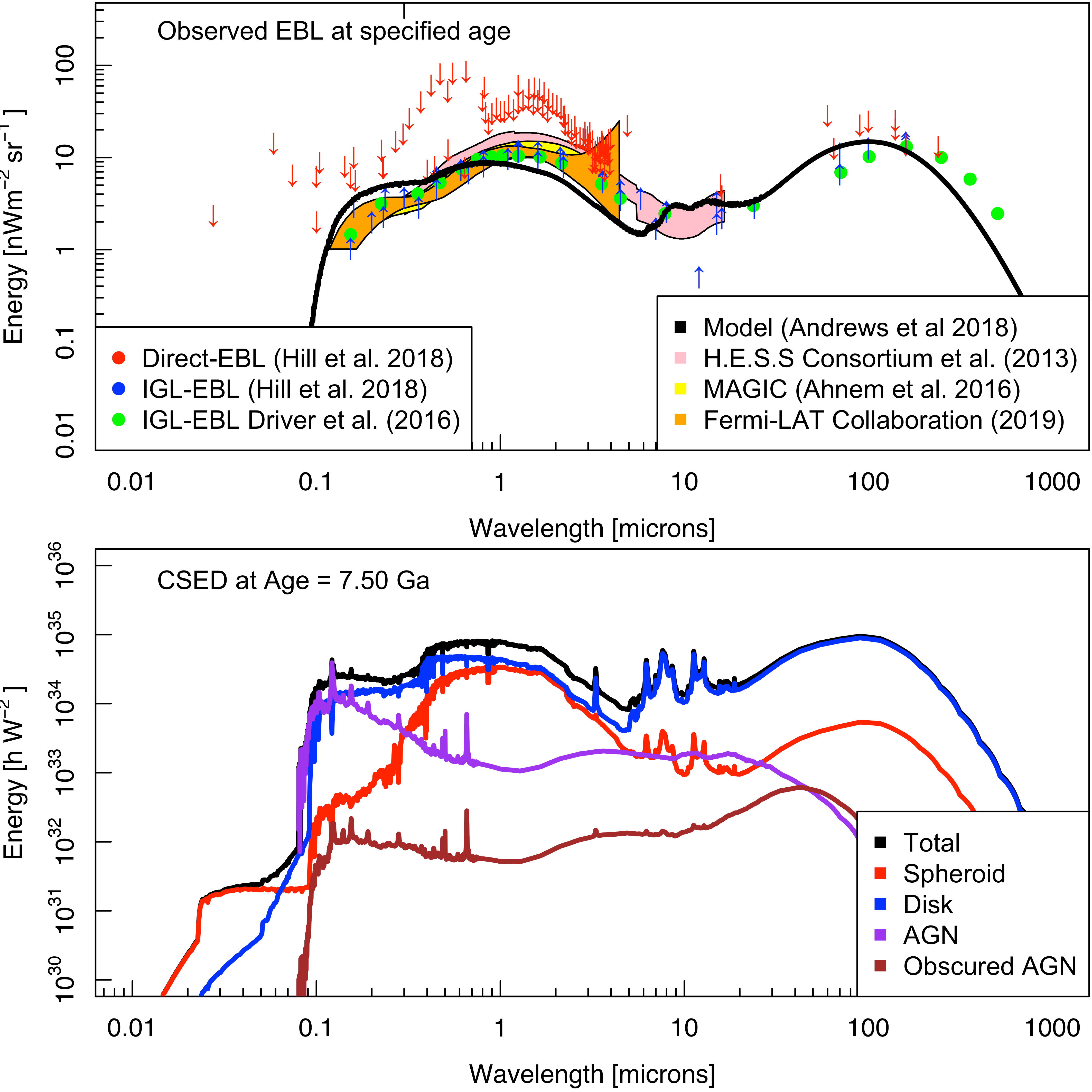}
 \caption{A static frame from our \href{https://www.dropbox.com/s/loxtb9svzr81tpr/iau355_driver_eblmovie.gif?raw=1}{EBL/CSED movie}, which shows
   the build up of the EBL over time. ({\it upper panel}) the EBL data
   as observed today with the EBL as it would be if observed when the
   Universe was 7.5Ga. ({\it lower panel}) the instantaneous cosmic
   spectral energy distribution (CSED) at an age of 7.5Ga, showing the
   contribution to the EBL from various components at this time-step
   (as indicated). The endpoint of the movie is the purple curve shown
   on Fig.\ref{fig:ebl}}
  \label{fig:models}
\end{center}
\end{figure}

\section{The state of EBL modelling: phenomenological and {\it apriori}}
In addition to SkySURF, there are a number of upcoming wide and deep
missions which will contribute significant new data in the coming
years. In particular high-resolution imaging data from new
space-platforms: Euclid, JWST, and Roman, and extremely wide
ground-based data from LSST. These should provide the statistical
power to reach that 1 per cent goal for the COB within a decade.

At the 1 per cent level interesting astrophysics arises and in
particular the direct contribution of reionisation becomes
quantifiable, as does the contribution from Intra Cluster Light (ICL),
Intra Group Light (IGL), Intra Halo Light (IHL), tidal streams, and a
myriad of other physical processes. 

At the present time there are a number of approaches to modelling the
EBL, which can be grouped under {\it apriori} models and
phenomenological models. Examples of the {\it apriori} approach arise
from numerical, semi-analytic and hydro-dynamical simulations such as
\cite{gilmore2012}, \cite{inoue2013}, \cite{cowley2019},
\cite{lagos2019}, and \cite{baes2019}. Examples of the
phenomenological model include our own work \cite{andrews2018,koushan2021}, along
with those of \cite{dominguez2011} and \cite{ks19}.

Here in brief we summarise our own model, described in full in
\cite{andrews2018}, which starts with two simple axioms: (1) the
formation of today's spheroid and bulge stars dominated the cosmic
star-formation history at high redshift, and (2) AGN growth and
activity is closely linked to spheroid and bulge star-formation. The
basis for the former is that the oldest known stars are found in the
Galactic Centre \citep{zoccali2006}, and for the latter we cite the
Gebhardt-Magorrian relation \citep{gebhardt2000,magorrian1998}.

In addition we require a few further decisions around the cosmic
star-formation history (see Fig.~\ref{fig:csfh}), an adopted stellar
population synthesis code, a dust attenuation model, and a metallicity
history. Here we adopt an invariant Chabrier IMF, the galaxy and AGN
dust attenuation models of \cite{dalehelou}, and a metallicity
evolution which linearly tracks the star-formation history. With these
two axioms and the choices above, we can produce the purple line shown
in Fig.~\ref{fig:ebl}. This maps the currently measured COB/CIB
portion of the EBL to within ~30 per cent accuracy, i.e., comparable
to the measurement error, across the entire optical to far-IR
wavelength range, with only some tweaking of the AGN component
required to match the UV data.

Essentially this provides a self-consistency test by which the adopted
Cosmic Star-formation History (see Fog.~\ref{fig:csfh}), under the
most simplistic assumptions, fully predicts the present day
EBL. Moreover the model not only predicts the EBL, but energy (photon)
production at any time over the past 13 billion years. This is
highlighted by the snapshot from our linked movie
(Fig.~\ref{fig:models}), which shows the EBL (upper) and Cosmic
Spectral Energy Distribution (CSED; lower) at a time when the Universe
was 7.5Ga. In the upper panel we show the redshift zero EBL
measurements, i.e., the endpoint to where the EBL will eventually grow
to match, and the EBL as it would be observed at an age of 7.5Ga,
i.e., in its fairly fledgling state. In the lower panel we show the
instantaneous energy being produced by the four components (as
indicated). The CSED, sub-divided into spheroid, disc obscured and
unobscured AGN contributions, is potentially far more powerful than
the EBL, as it dissects the EBL across time and providing a clear
falsifiable prediction as our measurements improve. In due course
comparisons between CSED measurements \citep{andrews2017} and models
\citep{andrews2018,baes2019,lagos2019} have the potential to constrain
many of the assumptions adopted in the model \citep[see, in particular,][]{koushan2021}.

\section{Prognosis and future directions}
The prognosis for the EBL and its subdivision into time slices are
remarkably good, with significant effort underway on a number of
fronts which will rapidly advance both the empirical measurements and
our capacity to model the data. Critical will be the complement of
upcoming deep and complete spectroscopic campaigns to allow for the
deconstruction of the EBL into the CSED. 

With the analysis of the VST KiDS, VISTA VIKING and HST SkySURF data
as they stand, we should be able to attain a $\sim 3$ per cent
measurement of the EBL from UV to near-IR wavelengths. In combination
with spectroscopic and photometric redshifts we will also be able to
measure the contributions of each time-interval to the EBL in coarse
billion year slices extending out to half the age of the Universe with
major high-density spectroscopic surveys like DEVILS
\citep{davies2018} and WAVES \citep{driver2019}, and beyond with ESO
MOONS and Subaru PFS.

Modelling this data will present both a challenge and an opportunity
for to the simulation community. The HST SkySURF project will also
likely lead to advances in our understanding of the two key
foregrounds, the Zodiacal Light and the Diffuse Galactic Light - both
of significant importance in the era of precision Astronomy. With
future facilities such as Messier, Euclid, JWST, Roman and LSST the
prospect of a 1 per cent constraint on the UV to mid-IR portion of the EBL becomes viable.

Very, briefly we return to the even more aspirational issue of a full
EBL analysis from $\gamma$-rays to radio waves. This too is likely to
be transformed over the next decade with various deep and wide radio
surveys allowing us to construct comparable IGL-EBL constraints over a
broad wavelength range from 20cm to 10m. eROSITA will also improve
upon previous measurements of the CXB, particularly at the soft x-ray
end where some discrepancies are seen (see Fig.~\ref{fig:ebl}).

However, perhaps the most exciting prospect, is the potential to also
construct CSED measurements into the x-ray and radio domains through
the stacking of x-ray and radio data at the locations of known
galaxies, or groups of galaxies, to directly measure the diffuse x-ray
and radio continuum contributions as a function of time.

We thank the organisers for a very enjoyable and
productive meeting, and look forward to continuing these discussions
over what looks to be a very busy and productive decade ahead.

\begin{discussion}

\discuss{A. Dom\'{\i}nguez}{What is the contribution of AGN to the
  EBL\,?}  \discuss{S. Driver}{In general the AGN are sub-dominant
  throughout, except in the UV and mid-infrared. However, we do have
  some unexplained flux in the ultraviolet which we currently {\it
    fix} by boosting the AGN a little. If this high UV flux really is
  due to AGN, then it is AGN which has kept the IGM ionised and still
  doing so today.}

\end{discussion}

\end{document}